\documentclass[11pt,letter]{article}

\usepackage{poma_style}

\usepackage{graphicx}
\usepackage{amsmath,amsfonts,bm}
\usepackage{courier}
\usepackage{siunitx}
\usepackage[colorlinks=true,urlcolor=blue,linkcolor=black,citecolor=black]{hyperref}
\usepackage{subcaption}

\begin{document}

\begin{center}
{\Large\textbf{Towards Physics-Informed Neural Networks for Stiff Guitar String Vibrations}}
\end{center}

Xinmeng Luan$^1$, Kensuke Okada$^2$, Ryoya Tabata$^2$, Gary Scavone$^1$

[1] Computational Acoustic Modeling Laboratory, Center for Interdisciplinary Research in Music Media and Technology, Schulich School of Music, McGill University, Montreal, Canada

[2] Department of Research \& Development, Yamaha Corporation, Hamamatsu, Shizuoka, Japan

\section*{abstract}
Modeling stiff string vibrations is challenging due to their dispersive and high-frequency characteristics. This study investigates the effectiveness of Physics-Informed Neural Networks (PINNs) in simulating the transverse vibration of a one-dimensional linear stiff string with sharp initial conditions induced by plucking. The governing Partial Differential Equation (PDE), along with the associated initial and boundary conditions, is incorporated directly into the loss function of the neural network. For the reference measurements, a wire-breaking experiment was performed to excite the string, and its vibration response was captured using a laser profiler. The comparison between experimental measurements, finite-difference time-domain (FDTD) simulations, and PINN-based simulations shows good overall agreement, highlighting the potential of PINNs for modeling stiff-string vibrations.

% 4-8-page page

\section{Introduction}
The simulation of plucked string instruments has long been a fundamental topic in musical acoustics, especially in the context of physics-based sound synthesis for instruments such as the guitar and harp. A variety of approaches have been proposed, including digital waveguide (DWG) methods, finite-difference time-domain (FDTD) schemes, and other numerical techniques \cite{bilbao2009numerical}, each offering different trade-offs between computational efficiency and physical accuracy.
Stiff string models exhibit dispersive behavior due to bending stiffness, leading to inharmonicity in
the vibration spectrum \cite{ducceschi2016linear}.  This introduces high-frequency components that are perceptually
important, but challenging to capture accurately in numerical simulations.

In recent years, deep learning has emerged as a powerful tool for modeling complex physical systems. In particular, Physics-Informed Neural Networks (PINNs) \cite{raissi2019physics} have gained attention for solving partial differential equations (PDEs) by embedding physical laws directly into the learning process. Within musical acoustics, PINNs have been explored for simulating instrument-related phenomena, such as nonlinear bow–string interactions 
\cite{luan2025physics} and acoustic wave propagation in tubes \cite{yokota2024physics}. They have also been applied to inverse problems, including the reconstruction of structural vibrations, for instance, in violin top plates using near-field acoustic holography techniques \cite{olivieri2021physics, luan2024complex, luan2025pinnsfd, luan2025physics_waspaa}, as well as in the reconstruction of acoustic radiation fields for tubes \cite{luan2025acoustic_tube}.
Applying PINNs to structural vibration problems remains challenging, particularly for systems involving stiffness and dispersive effects. Recent studies on cable vibrations considering bending stiffness have demonstrated that classical PINNs often fail to accurately reproduce such dynamics,
highlighting the difficulty of capturing multi-scale behavior governed by higher-order derivatives \cite{dan2025pinn_cable}.

In this paper, we investigate the performance of PINNs for simulating the vibration of a plucked string. We
focus on a linear stiff string model under the assumption of small-amplitude vibrations. 
%This setting provides a particularly challenging benchmark for PINNs, due to the combination of dispersive dynamics and sharp initial conditions. The results highlight that a key challenge in this setting lies in the sharp initial conditions introduced by plucking, which may be difficult for neural networks to accurately represent due to their inherent spectral bias toward smooth functions. This is particularly relevant for stiff string models, where dispersion causes each frequency component to propagate at a different speed, resulting in complex multi-scale dynamics that are difficult to represent accurately. To address this issue, we employ a Fourier feature embedding, which has been shown to improve the representation of high-frequency components in neural networks. 
This setting is particularly challenging for PINNs because it combines sharp pluck-induced initial conditions with dispersive stiff-string dynamics. The pluck introduces a non-smooth initial profile containing substantial high-frequency content, which is difficult for neural networks to represent accurately due to their spectral bias toward low-frequency functions. In a stiff string, the bending stiffness term introduces dispersion, causing higher-frequency components to propagate faster than lower-frequency ones. This effect becomes significant precisely in the high-wavenumber regime where spectral bias is most severe. To address this, we employ a Random Fourier Feature embedding (RFF), which mitigates spectral bias and thereby provides the network with the capacity needed to resolve the physically correct dispersion relation across a broader range of wavenumbers.
This work provides a preliminary study in a physically relevant setting, supported by experimental validation, aimed at assessing the suitability of PINNs for plucked string applications. 
%In addition, experimental measurements are incorporated to support the analysis.
The insights gained from this study are intended to inform future developments, including inverse problems such as parameter estimation, as well as extensions to more advanced frameworks such as physics-informed neural operators.

\section{Plucked string vibration}
Consider a string of length $L$ and cross-sectional area $A$, linear density $\rho$, subject to a constant tensile force $T$, which governs its transverse vibration. 
 In addition, the string exhibits bending stiffness associated with longitudinal elastic deformation, characterized by Young’s modulus $E$ and the moment of inertia $I$, introducing dispersive effects into the vibration.
Dissipation is incorporated following \cite{bensa2003simulation} through a composite damping model consisting of a frequency-independent term $\sigma_0$  and a frequency-dependent term $\sigma_1$.
Under the assumption of small-amplitude vibrations, the transverse displacement of the string, denoted by  $u(x,t)$  as a function of spatial coordinate along the axial direction $x$ and time $t$, is governed by the following PDE:

\begin{equation}
    \rho A \partial_t^2u - T \partial_x^2u + EI\partial_x^4u + 2\rho A \sigma_0 \partial_t u - 2 \rho A \sigma_1 \partial_t \partial_x^2 u =0, \label{eq:pde_or}
\end{equation} 
where $\partial_x$ and $\partial_t$ represent partial derivatives with respect to $x$ and $t$. This equation has been widely employed in the field of musical acoustics \cite{ducceschi2016linear}.
 Here, geometric nonlinearities associated with large-amplitude motion are neglected. Nevertheless, this simplification does not preclude the adoption of the same computational framework for extension to such nonlinear model \cite{bilbao2005conservative}.

%$\rho$ the material density in \si{\kilo\gram\per\meter\tothe{3}}, $A= \pi r^2$ in \si{\meter \squared} the string cross-sectional area for a string radius $r$ in \si{\meter}, T the tension in \si{\newton}, $E$ in \si{\newton\meter}, $I$ in \si{\meter\tothe{4}} the moment of inertia.

\begin{figure}
    \centering
    \includegraphics[width=0.5\linewidth]{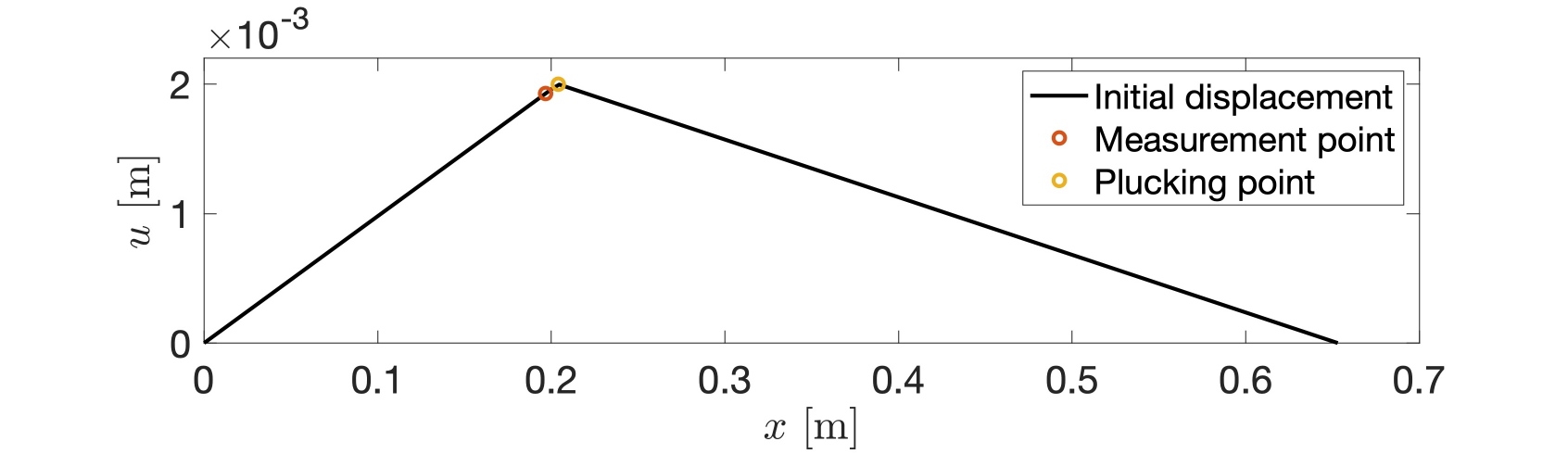}
    \caption{Initial condition for plucked string vibration.}
    \label{fig:ic}
\end{figure}

A simple plucked-string condition is adopted following Perov et al. \cite{perov2016physics}, where a triangular displacement profile (see Fig.~\ref{fig:ic}) is used as the initial condition (ICs):
\begin{equation}
\begin{aligned}
&u(x,0) =
\begin{cases}
\frac{y}{x_1}\, x, & 0 \leq x \leq x_1, \\
\frac{y}{L - x_1}\, (L - x), & x_1 \leq x \leq L,
\end{cases} \\
& \partial_t u(x,0) = 0,
\end{aligned} 
\label{eq:ic}
\end{equation}
where the apex of the triangular profile is located at $x_1$, and the displacement reaches its maximum value $y$ at this point.

The following Boundary Conditions (BCs) is considered:
\begin{equation}
	u(0,t) = u(L,t) = 0, \quad 
	\partial_x^2u(0,t) = \partial_x^2u(L,t) = 0.
	\label{eq:bc}
\end{equation}

\section{Physics-Informed Neural Networks}

%soft/hard constraints

%RFF

The spatial and temporal coordinates  $x$ and $t$  are used as inputs to the PINN. To improve training stability and performance, these variables are typically normalized. In addition, reducing the number of independent physical parameters is beneficial for both efficiency and generalization. Therefore,  $x$ and $t$  are non-dimensionalized using $L$ and the period $T_p$, respectively:
\begin{equation}
\begin{aligned}
x' &= \frac{x}{L} \in [0,1], \quad
t' = \frac{t}{T_p} \in [0,1], \\
\kappa' &= \frac{2}{L} \sqrt{\frac{EI}{T}}, \quad
\sigma_0' = \frac{2}{L} \sigma_0 \sqrt{\frac{\rho A}{T}}, \quad
\sigma_1' = \frac{2}{L} \sigma_1 \sqrt{\frac{\rho A}{T}}.
\end{aligned}
\end{equation}
This transformation reduces the original set of physical parameters
$\{L, \rho, A, T, E, I, \sigma_0, \sigma_1\}$
to a more compact set of dimensionless parameters
$\{\kappa', \sigma_0', \sigma_1'\}$. 
Subsequently, the PDE, ICs, and BCs are consistently normalized; however, their explicit forms are omitted for brevity and introduced later in the loss function.

A deep neural network is employed, with $x'$ and $t'$ as inputs, and $\hat{u}(x',t')$ as the predicted output:
\begin{equation}
   \hat{u} (x',t') = \Theta_{\theta} ( \Gamma (x', t')),
   \label{eq: phi_formu}
\end{equation}
where $\theta_\theta$ denotes the neural network-based estimator, parameterized by the trainable parameters $\theta$. $\Gamma$ denotes a RFF embedding \cite{rahaman2019spectral}, which has been shown to effectively mitigate the spectral bias in PINNs \cite{wang2021eigenvector}. Spectral bias refers to the tendency of neural networks to preferentially learn low-frequency components before high-frequency ones, often leading to difficulties in accurately representing high-frequency phenomena.
The RFF is defined as
\begin{equation}
\Gamma(\mathbf{v}) =
\big[ \cos(\omega_1^\top \mathbf{v}), \sin(\omega_1^\top \mathbf{v}), \ldots,
\cos(\omega_i^\top \mathbf{v}), \sin(\omega_i^\top \mathbf{v}), \ldots,
\cos(\omega_M^\top \mathbf{v}), \sin(\omega_M^\top \mathbf{v})\big],
\end{equation}
where $\mathbf{v}=(x',t')$ and the frequency vectors are independently sampled from a Gaussian distribution,
$\omega_i \sim \mathcal{N}(0, \sigma^2)$.

Automatic differentiation is then used to compute the loss functions, which encompass the PDE, ICs and BCs losses, all formulated as mean squared error terms. 
The PDE loss is expressed as according to \eqref{eq:pde_or}
\begin{equation}
    \mathcal{L}_{PDE} = \frac{1}{N_{PDE}}    \big \|   
    \partial_{t'}^2\hat{u} - 4 \partial_{x'}^2\hat{u}+ \kappa '^2\partial_{x'}^4\hat{u} + 2\sigma_0 ' \partial_{t'} \hat{u} - 2  \sigma_1 ' \partial_{t'} \partial_{x'}^2 \hat{u}
    \big \|_2^2.
\end{equation}
The ICs loss is express as according to \eqref{eq:ic}
\begin{equation}
\begin{cases} \displaystyle{
     \mathcal{L}_{IC1} = \frac{1}{N_{IC1}} \big \| \hat{u}(x',0) - {u}\big (x',0) \big \|_2^2,} \\ 
     \displaystyle{
     \mathcal{L}_{IC2} = \frac{1}{N_{IC2}} \big \| \partial_t \hat{u}(x',0) - \partial_t {u}\big (x',0) \big \|_2^2.}  
\end{cases}
\end{equation}
The BCs loss is express as according to \eqref{eq:bc}
%\begin{equation}
%    \mathcal{L}_{BC} = \frac{1}{N_{BC}} \big \| \hat{u} - u \big \|_2^2, \quad x =0, \quad t \in [0,T].
%\end{equation}
\begin{equation}
\begin{cases} \displaystyle{
     \mathcal{L}_{BC1} = \frac{1}{N_{BC1}} \big \| \hat{u}(0,t') - {u}\big (0,t') \big \|_2^2,} 
     \\ 
     \displaystyle{
     \mathcal{L}_{BC2} = \frac{1}{N_{BC2}} \big \| \hat{u}(1,t') - {u}\big (1,t') \big \|_2^2,} 
     \\ 
     \displaystyle{
     \mathcal{L}_{BC3} = \frac{1}{N_{BC3}} \big \| \partial_x^2 \hat{u}(0,t') - \partial_x^2 {u}\big (0,t') \big \|_2^2,}  
          \\ 
     \displaystyle{
     \mathcal{L}_{BC4} = \frac{1}{N_{BC4}} \big \| \partial_x^2 \hat{u}(1,t') - \partial_x^2 {u}\big (1,t') \big \|_2^2.}  
\end{cases}
\end{equation}
$N_{PDE}$, $N_{IC1}$, $N_{IC2}$, $N_{BC1}$,  $N_{BC2}$, $N_{BC3}$ and  $N_{BC4}$ represent the respective numbers of collocation points used for the loss computations.
Then the total loss function is 
\begin{equation}
\begin{aligned}
    \mathcal{L} &= \lambda_{PDE} \mathcal{L}_{PDE} + \lambda_{IC1} \mathcal{L}_{IC1} + \lambda_{IC2} \mathcal{L}_{IC2}  + \lambda_{BC1} \mathcal{L}_{BC1} + \lambda_{BC2} \mathcal{L}_{BC2}  + \lambda_{BC3} \mathcal{L}_{BC3}  + \lambda_{BC4} \mathcal{L}_{BC4} ,
\end{aligned}
\end{equation}
with $\lambda_{PDE} $, $\lambda_{IC1}$, $ \lambda_{IC2}$, $\lambda_{BC1}$, $ \lambda_{BC2}$, $\lambda_{BC3}$, $ \lambda_{BC4}$,  as the loss function weights. Modern optimization algorithms are employed to iteratively update the network parameters $\theta$ during training.
\section{Validation}

\subsection{PINN Implementation}

The neural network architecture is based on a multilayer perceptron (MLP), with 2 input channels corresponding to $(x,t)$, a single output channel $u$, and 4 hidden layers of width 256. A hyperparameter study is conducted using 4 PINN models with $\sigma \in \{1, 5, 10, 15\}$ for RFF, where the RFF feature dimension is set to 256. The $\tanh$ activation function is adopted throughout. The MLP output is scaled by the maximum value of the initial displacement.

For training, we employ the state-of-the-art second-order optimizer Shampoo with Adam in the preconditioner’s eigenbasis (SOAP) \cite{vyas2024soap}. SOAP has been shown to effectively approximate the Hessian-based preconditioner, resulting in notable performance gains in PINN training \cite{wang2025gradient}.
Moreover, as continuous-time PINNs may violate causality and converge to physically inconsistent solutions \cite{wang2024respecting}, we adopt a causal training strategy to mitigate time-related issues. We also incorporate the GradNorm method \cite{chen2018gradnorm} to adaptively balance the loss terms, avoiding manual tuning of loss weights. All implementations are carried out using JAX. As a preliminary study, the temporal domain is restricted to the first vibration period. Nevertheless, the proposed approach is not limited to short durations and can be extended to longer time intervals by increasing the number of PINNs, as demonstrated in Luan et al. \cite{luan2025physics}

\subsection{Experiments}

\begin{figure}[h!]
    \centering
    \includegraphics[width=0.3\linewidth]{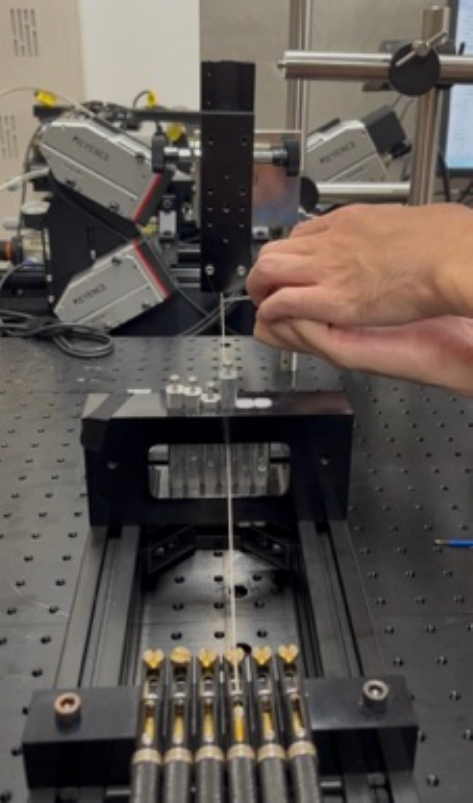}
    \caption{Custom-built wire-breaking system.}
    \label{fig:wirebreak}
\end{figure}

In this study, a wire-breaking method is employed using a custom-built experimental system (see Fig.~\ref{fig:wirebreak}). The excitation is generated by suddenly releasing a pre-tensioned string, which is cut using a pair of scissors to approximate an ideal impulsive release. This approach produces an initial condition dominated by transverse displacement in a single direction, with negligible initial velocity, closely resembling a pluck-like excitation while maintaining good repeatability.

%\footnote{\href{https://www.daddario.com/collections/electric-guitar-strings/products/exl110-xl-nickel-wound-electric-guitar-strings-regular-light-10-46}{https://www.daddario.com/collections/electric-guitar-strings/products/exl110-xl-nickel-wound-electric-guitar-strings-regular-light-10-46}}
The experiments are conducted using a D'Addario low E electric guitar string ($82.4~\mathrm{Hz}$) \cite{daddario}.
The string parameters are given as: density $\rho = 6448~\mathrm{kg/m^3}$, diameter $d = 1.18~\mathrm{mm}$, Young’s modulus $E = 200~\mathrm{GPa}$, length $L = 0.653~\mathrm{m}$, and tension $T = 80.42~\mathrm{N}$. The damping coefficients are set to $\sigma_0 = 0.1$ and $\sigma_1 = 1 \times 10^{-5}$, with excitation applied at position $x_1 = 0.204~\mathrm{m}$, initial displacement amplitude $y = 2 \times 10^{-3}~\mathrm{m}$, and measurement point at $0.2~\mathrm{m}$.
 The resulting transverse vibration is measured using a high-speed 2D laser profiler \cite{keyence_ljv7000}, which is capable of capturing both polarization components. The measurement sampling frequency is $2560~\mathrm{Hz}$. However, in this work, only a single transverse direction is considered for analysis to simplify the modeling framework and focus on the dominant vibration behavior.
%\footnote{\href{https://www.keyence.com/products/measure/laser-2d/lj-v/}{https://www.keyence.com/products/measure/laser-2d/lj-v/}}

\subsection{Results and discussion}

%The selection of RFF can be based on audio rate. the connection between audio and rff. 
%A plot of frequency domain loss.

\begin{figure}[t]
    \centering
    \begin{subfigure}{0.19\linewidth}
        \centering
        \includegraphics[width=\linewidth]{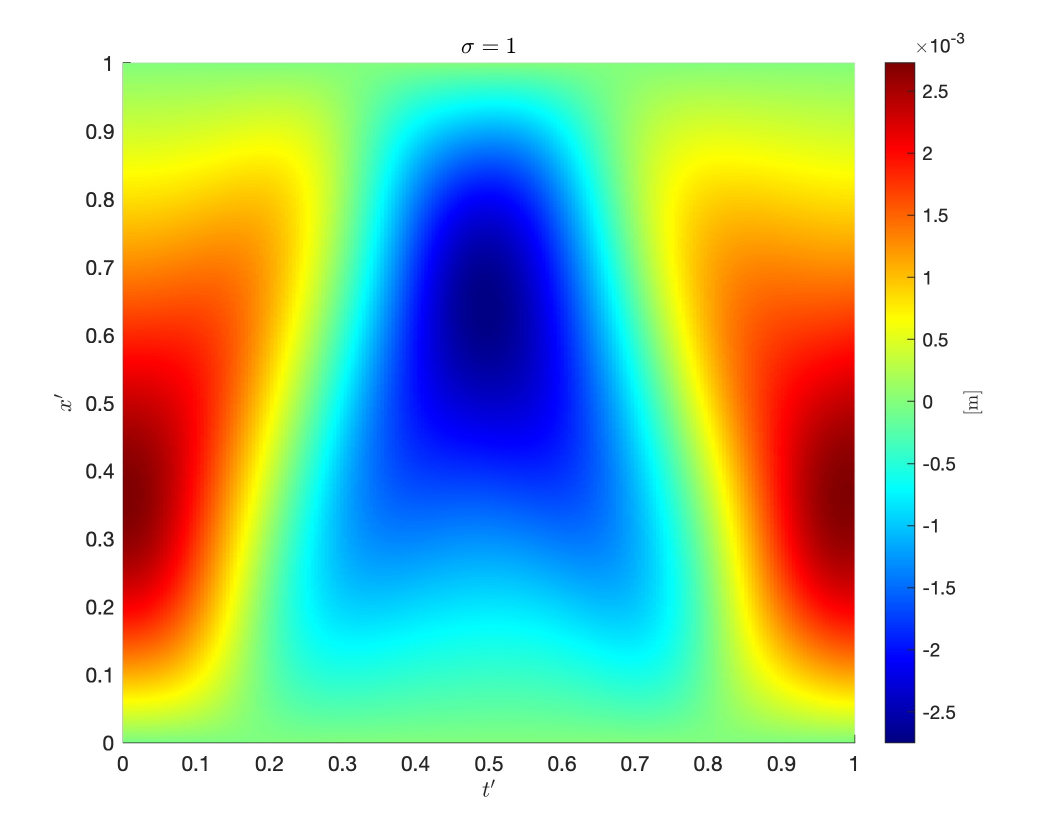}
        % \caption{Caption}
    \end{subfigure}
    \hfill
    \begin{subfigure}{0.19\linewidth}
        \centering
        \includegraphics[width=\linewidth]{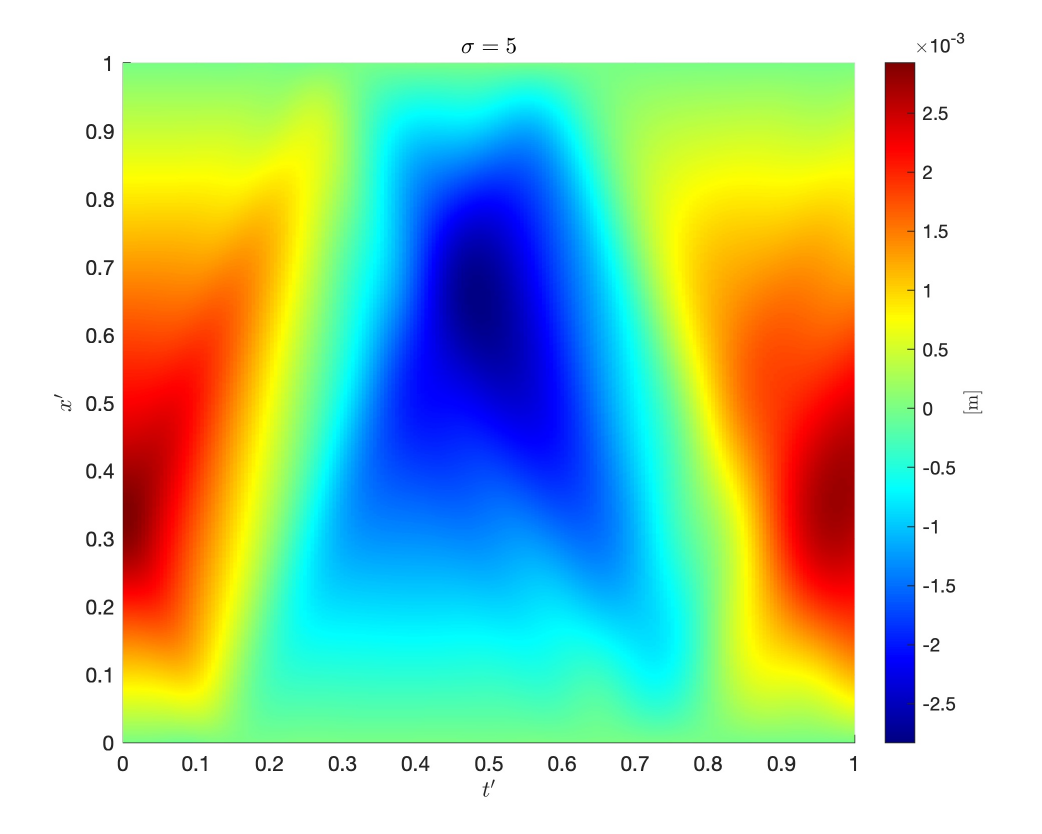}
        % \caption{Caption}
    \end{subfigure}
        \hfill
    \begin{subfigure}{0.19\linewidth}
        \centering
        \includegraphics[width=\linewidth]{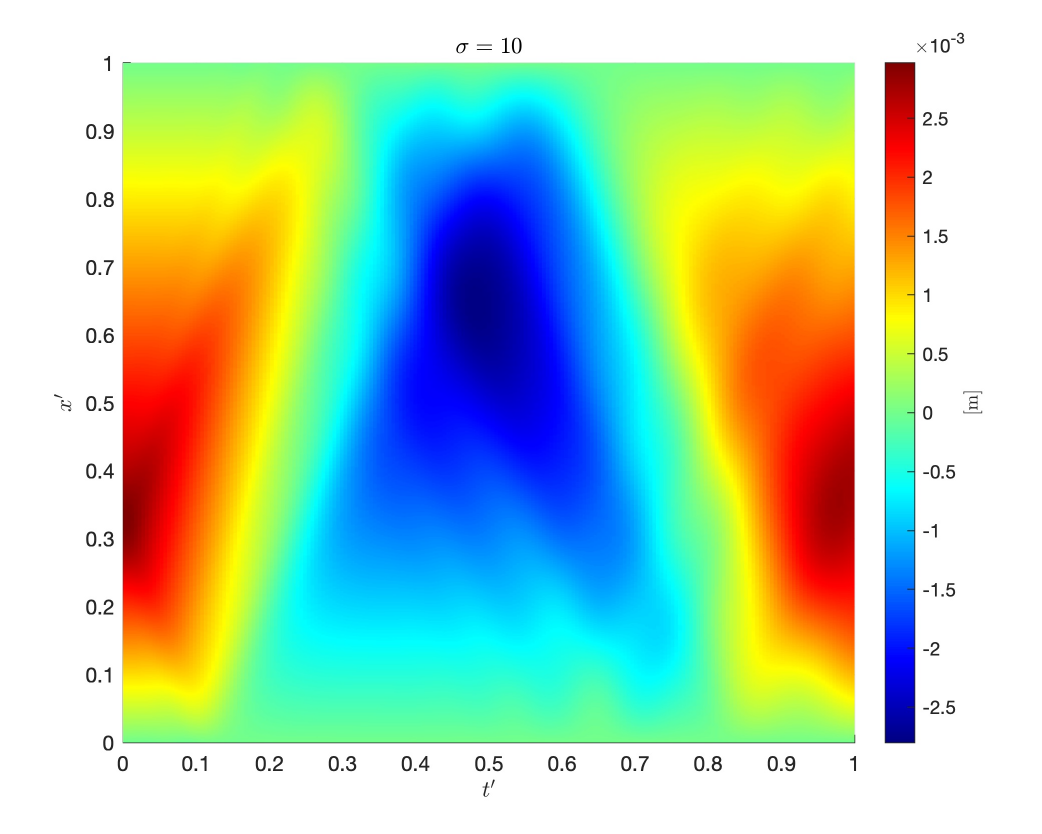}
        % \caption{Caption}
    \end{subfigure}
        \hfill
    \begin{subfigure}{0.19\linewidth}
        \centering
        \includegraphics[width=\linewidth]{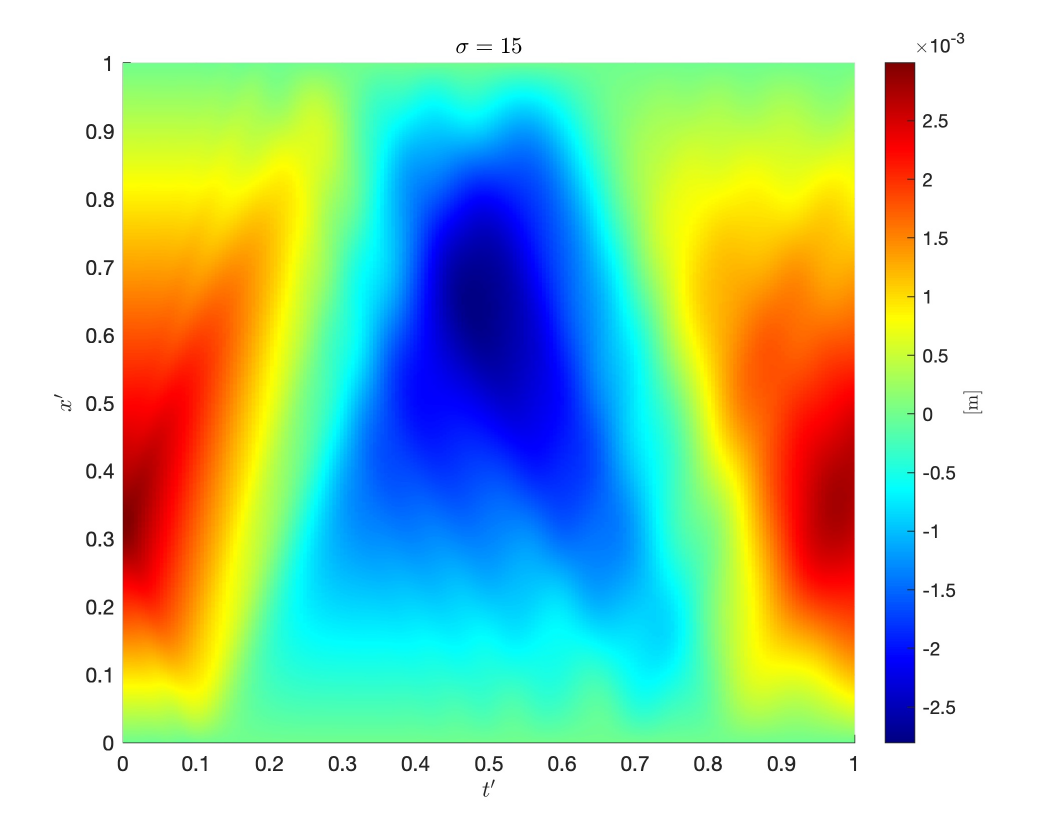}
        % \caption{Caption}
    \end{subfigure}
        \hfill
    \begin{subfigure}{0.19\linewidth}
        \centering
        \includegraphics[width=\linewidth]{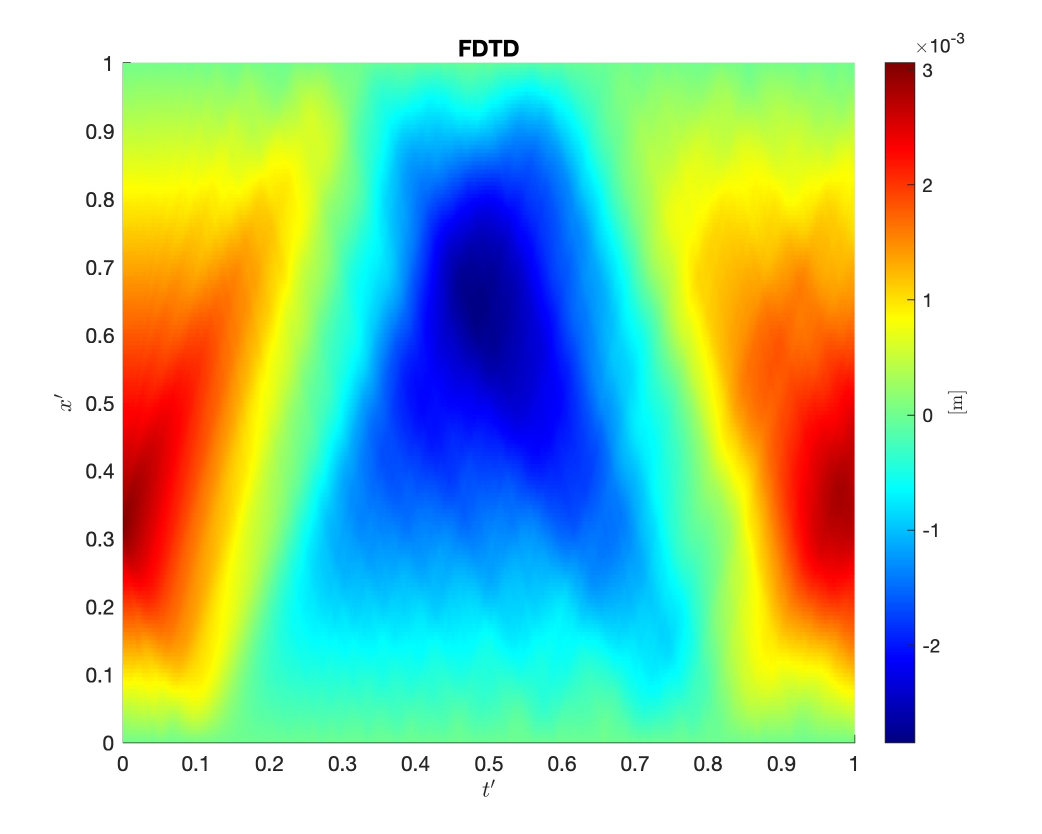}
        % \caption{Caption}
    \end{subfigure}
    \caption{Comparison of $u(x,t)$. From left to right: PINN predictions with $\sigma = 1, 5, 10, 15$, and the FDTD solution.}
        \label{fig:rff_combined}
\end{figure}

Figure~\ref{fig:rff_combined} presents the PINN simulation results of the field $u(x,t)$ under different RFF scales $\sigma$, alongside the corresponding FDTD baseline for comparison. When $\sigma = 15$, the result closely matches the FDTD solution, demonstrating the ability of PINNs to accurately model plucked string vibrations.
Moreover, it is observed that larger values of $\sigma$ yield improved results, with enhanced fine-scale details and higher resolution. 
The Mean Square Error (MSE) between the PINN predictions and the FDTD reference is summarized in Table~\ref{tab:mse_sigma}, where $\sigma = 15$ achieves the best performance.
\begin{table}[h]
\centering
\caption{MSE between PINN and FDTD for different values of $\sigma$.}
\label{tab:mse_sigma}
\begin{tabular}{c c}
\hline
$\sigma$ & MSE \\
\hline
1  & $9.7 \times 10^{-9}$ \\
5  & $3.4 \times 10^{-9}$ \\
10 & $3.3 \times 10^{-9}$ \\
15 & $2.7\times 10^{-9}$ \\
\hline
\end{tabular}
\end{table}

In audio applications, the human auditory range extends up to approximately 22 kHz, making accurate high-frequency representation essential. The results indicate that increasing $\sigma$ in the RFF can significantly enhance the model’s ability to capture high-frequency components, consistent with the expectation reported in Rahaman et al. \cite{rahaman2019spectral}.
The issue of insufficient high-frequency content in PINNs has also been reported in tube acoustics modeling \cite{yokota2024physics}. 

\begin{figure}[t]
    \centering
    \begin{subfigure}{0.48\linewidth}
        \centering
        \includegraphics[width=\linewidth]{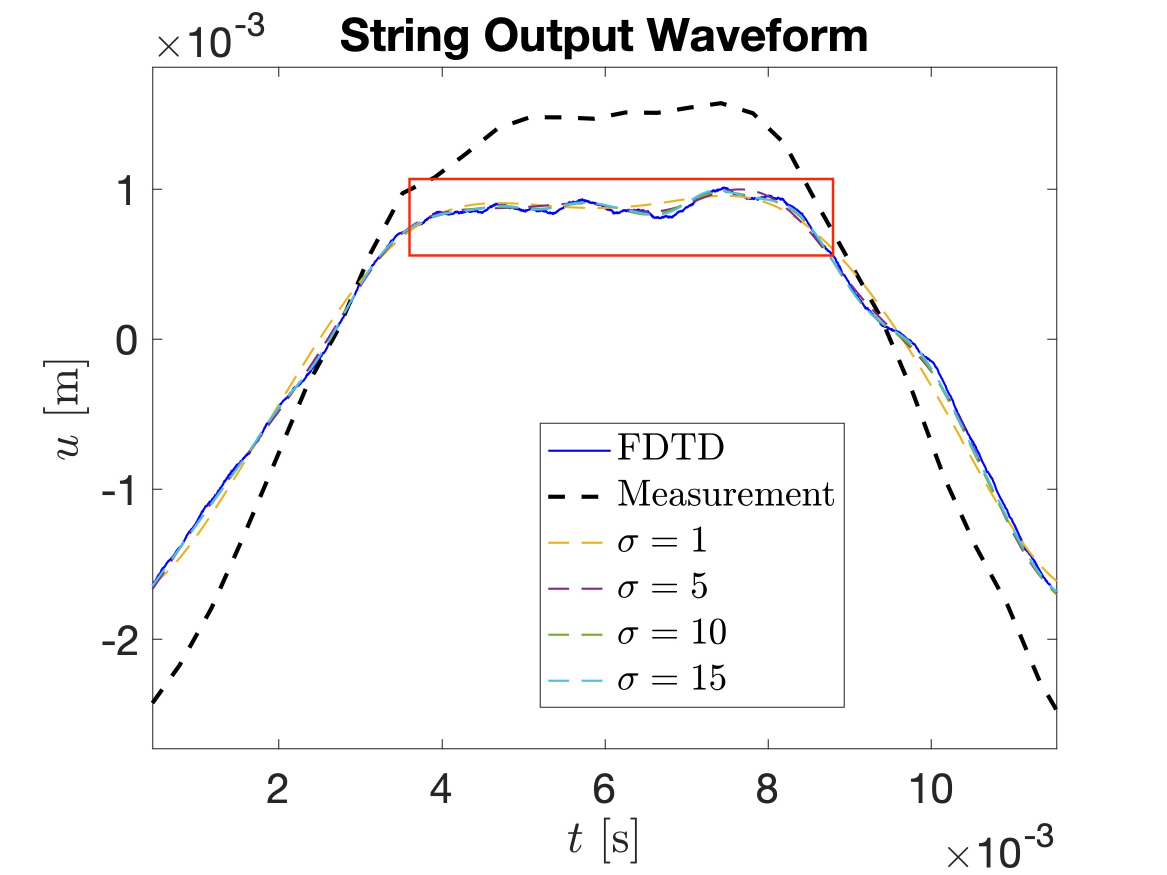}
        \caption{In the first period.}
        \label{fig:rff}
    \end{subfigure}
    \hfill
    \begin{subfigure}{0.48\linewidth}
        \centering
        \includegraphics[width=\linewidth]{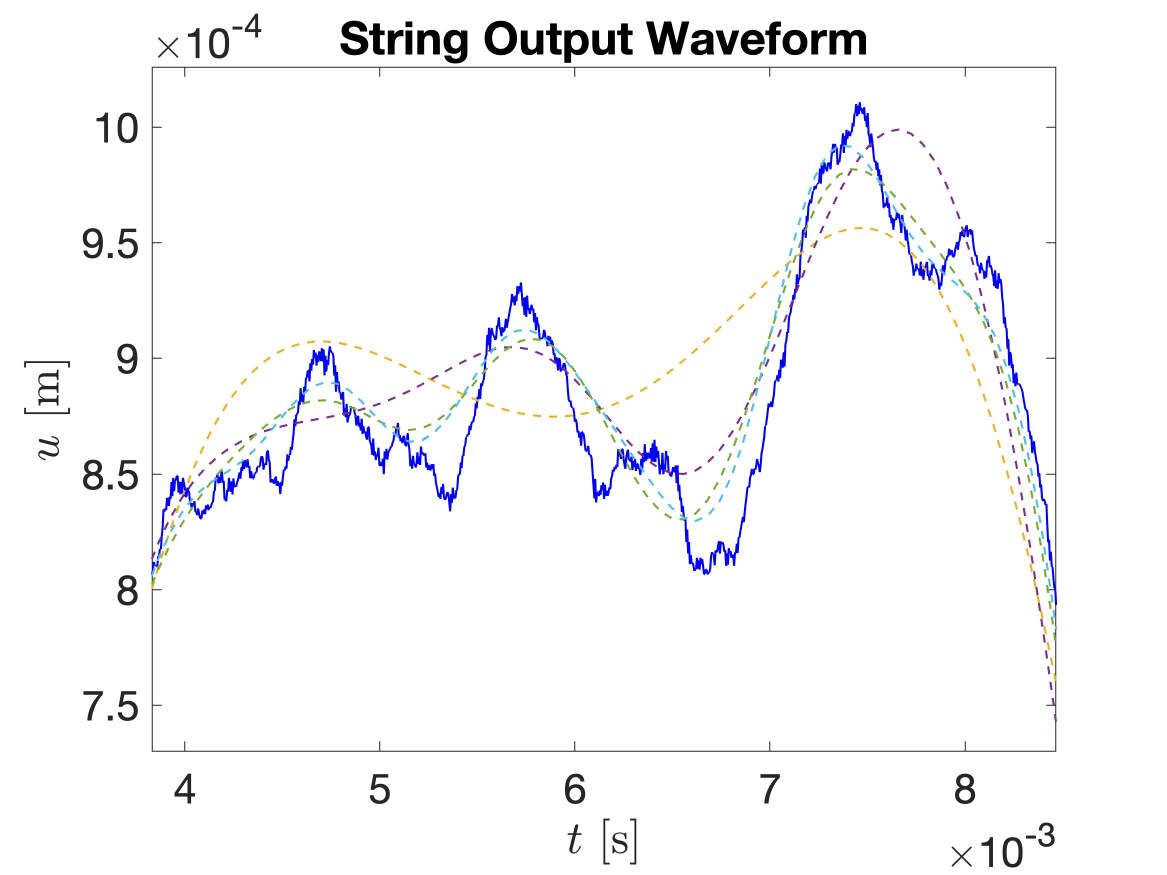}
        \caption{Zoomed-in view of the red boxed region in the left figure.}
        \label{fig:rff_zoom}
    \end{subfigure}
    \caption{Comparison of string displacement at the laser measurement point: experimental data, PINN predictions (with varying RFF $\sigma$), and FDTD results.}
    \label{fig:meas}
\end{figure}

A visualization of the measured string vibration, together with simulation results from PINNs (with different RFF $\sigma$ values) and FDTD, is presented in Fig.~\ref{fig:meas}. Comparing the PINN and FDTD results, particularly in the zoomed view on the right, shows that larger $\sigma$ values enable the model to capture higher-frequency content more effectively. When comparing the simulations to the measurements, a noticeable amplitude discrepancy is observed. This difference may arise from experimental limitations, as it is difficult to precisely control the initial waveform during measurement, and an ideal triangular excitation is not perfectly realized in practice. In addition, the sampling frequency of the laser measurements is lower than that of the simulations, resulting in a smoother waveform with reduced high-frequency content.

Overall, the comparison results from this preliminary analysis indicate that PINNs offer a promising approach for modeling string vibrations. A key limitation is that only the first period of the signal is evaluated, which is clearly too short; however, this serves as an initial assessment of whether PINNs can achieve reasonable performance for the considered PDE. We emphasize that a central challenge in applying PINNs to plucked string simulation is accurately capturing the high-frequency content. It is worth noting that the fourth-order derivative in \eqref{eq:pde_or} does not hinder training, which is encouraging, given that higher-order PDEs generally exhibit a lower probability of stable gradient flow convergence \cite{song2024does}.
By construction, PINNs represent a dynamical system under fixed ICs and BCs through a parameterized neural network. Owing to this formulation, the resulting model is fully differentiable, making it directly amenable to gradient-based optimization methods. This property enables its use in inverse problems, including parameter estimation from measurements and sound resynthesis, and warrants further investigation.

\section{Conclusion}

This study demonstrates that PINNs can effectively simulate the transverse vibration of a stiff string, showing good agreement with both experimental measurements and FDTD results. Although the analysis is limited to the first vibration period, it provides an initial validation of the approach. Capturing high-frequency content remains a key challenge, but the inclusion of a fourth-order derivative does not hinder training. Owing to their fully differentiable formulation, PINNs also offer strong potential for inverse problems, motivating further investigation in musical acoustics applications. Future work will investigate extensions to nonlinear string models.

% \section*{Acknowledgments}

% Acknowledgments, if any, follow the conclusion section.

% \appendix
% \section*{Appendix A}

% Appendices, if necessary, can go here. However, given the scope of a typical POMA article, the use of an appendix would be uncommon.

% \begin{thebibliography}{9}

% \bibitem{daigle79}

% \end{thebibliography}

\bibliographystyle{IEEEtran} 
\bibliography{sampbib.bib}

\end{document}